# First CT-MRI Scanner for Multi-dimensional Synchrony and Multi-physical Coupling
– A Major Step towards the Grand Fusion "*Omni-tomography*"
("*All-in-One*" & "*All-at-Once*")

Ge Wang, PhD (wangg6@rpi.edu, http://www.rpi-bic.org), 02/10/14

**Abstract:** We propose to prototype the first CT-MRI scanner for radiation therapy and basic research, demonstrate its transformative biomedical potential, and initiate a paradigm shift in multimodality imaging. Our design consists of a double donut-shaped pair of permanent magnets to form a regionally uniform ~0.5T magnetic field and leave room for a stationary 9-source interior CT gantry at 3 tube voltages (triple-energy CT). Image reconstruction will be in a compressive sensing framework. Please discuss with Dr. Ge Wang (ge-wang@ieee.org) if you are interested in collaborative opportunities.

### A. Overview

We propose to prototype the first CT-MRI scanner, demonstrate its transformative biomedical potential, and initiate a paradigm shift in multimodality imaging. <u>First, the project is technically innovative</u>. Such a scanner has not been attempted before, because both CT and MRI scanners are bulky, and a CT scanner has fast spinning heavy metals (source and detectors, ~3 turns per second), while an MRI scanner has a great magnetic field measured in Tesla (25-65 micro Tesla for the geomagnetic field). Thus, a rotating gantry is prohibitive in a high magnetic field. Fortunately, it is interior tomography we pioneered [1] that offers an unconventional way to combine CT and MRI, and beyond. <u>Second, the project is clinically important</u>. CT is best at structural details and rapid scanning, while MRI is superior at soft-tissue contrast and functional imaging. With its inherent co-registration in space, time and spectra, the CT-MRI scanner has major utilities for cardiac imaging (CT angiography fused with MRI perfusion), cancer imaging (contrast-enhanced heterogeneity characterization), imaging-guided surgical intervention and radiation therapy. <u>Third, the project is potentially revolutionary</u>. With the CT-MRI scanner, we can extract not only spatiotemporal correlation of complicated features but also <u>multi-physics coupling based parameters</u>. We are working on possibilities to integrate nanoparticles, x-ray imaging, and MRI (NXMRI). A key idea is that x-rays modulate distributions of electrons inside nanoparticles, and change MRI parameters such as T2. Synergy from synchrony and coupling could be remarkable for CT-MRI to accelerate the development of systems biomedicine.

### B. Research Plan

In collaboration with partners, we plan to prove the concept that CT-MRI can (1) <u>improve radiation therapy</u> and (2) <u>enable basic research</u>. To suppress interference between CT and MRI, we will use a stationary CT architecture in which multiple sources are fixed around a patient, along with the corresponding detectors. All the x-ray beams can focus on a region of interest (ROI), representing a few-view interior imaging setup. Since the CT components are mechanically motionless during operation, the electromagnetic shielding for CT and MRI can be greatly simplified. As the patient posture remains the same, correlation between CT and MRI features will be utilized to reduce both CT and MRI datasets. More specifically, our design consists of a double donut-shaped pair of permanent magnets to form a regionally uniform ~0.5T magnetic field and leave room for a stationary 9-source interior CT gantry at 3 tube voltages (triple-energy CT). Image reconstruction will be in a compressive sensing framework.

The first example will use a large animal model to demonstrate that CT-MRI provides much richer information on cancer than either CT or MRI alone, and better define tumor margins in real-time, and improve therapeutic outcomes [2].

As the second example also with a large animal model, interactions among osteoporotic mechanisms, novel nanoparticles, x-rays and RF waves will be observed and manipulated by the CT-MRI prototype for unique insights (x-ray excited nanoparticles would show different MR behaviors corresponding to various physiological and pharmaceutical states).

While the second example is out-of-box and aggressive, the first example is clinically important and clearly translational (the state-of-the-art of radiation therapy is LINAC-MRI [3], which suffers from deformed geometry of on-board MRI and approximated attenuation with prior CT).

The CT-MRI system will be developed in the Biomedical Imaging Center, to be a highlight of the new imaging corridor within the Center for Biotechnology and Interdisciplinary Studies (CBIS). This project will not only strengthen our external partnerships but also promote campus-wide collaborative efforts within CBIS and with other groups on campus. RPI has an umbrella license from Virginia Tech to work on CT-MRI, and strategic agreements with GE GRC and MSSM to facilitate joint biomedical imaging research.

## C. Perspective

We envision that tomography will transcend current modality fusion towards what we call "*the grand fusion*" of all relevant modalities, namely, "*omni-tomography*", for truly simultaneous but often localized image reconstruction in terms of many contrast mechanisms such as CT, MRI, PET, SPECT, and more [4]. The challenge in fusing modalities for simultaneous imaging has been *space conflict* and *other physical constraints*. Traditionally, scanners of different types are longitudinally assembled but this arrangement cannot synchronize data collection. In contrast, the proposed CT-MRI project is a major step towards the holy grail of biomedical imaging – omni-tomography. Over the past years, we have established interior tomography as a general tomographic principle, and applied it for interior SPECT, interior MRI, interior phase-contrast tomography, and so on. Therefore, relevant data acquisition modules can be rather compact to provide space and allow harmony for omni-tomography [5]. Based on the theoretical and technical foundation we have built, we are motivated to seek support of *just a few million dollars*, showcase a dreamed machine, and generate a huge and lasting impact on biomedicine.

## References

[1] http://iopscience.iop.org/0031-9155/58/16/R161

[2] https://spie.org/x94063.xml

[3] http://medicalphysicsweb.org/cws/article/research/56179

[4] http://www.plosone.org/article/info%3Adoi%2F10.1371%2Fjournal.pone.0039700

[5] http://medicalphysicsweb.org/cws/article/opinion/51026